\documentclass[twocolumn,showpacs,preprintnumbers,amsmath,amssymb]{revtex4}
\usepackage{graphicx}% Include figure files
\usepackage{dcolumn}% Align table columns on decimal point
\usepackage{bm}% bold math

\begin{document}

%\preprint{APS/123-QED}

\title{Continuous or discrete attractors in neural circuits? A self-organized switch at maximal entropy}% Force line breaks with \\

\author{Alberto Bernacchia}
%\email{a.bernacchia@gmail.com}
%\altaffiliation{}%Lines break automatically or can be forced with \\
\affiliation{Universita' ``La Sapienza'', Dipartimento di Fisica ``E.Fermi'', Roma%\\
}

%\affiliation{Second institution and/or address}

\date{\today}% It is always \today, today,

\begin{abstract}
A recent experiment suggests that neural circuits may alternatively implement 
continuous or discrete attractors, depending on the training set up. 
In recurrent neural network models, continuous and discrete attractors are separately 
modeled by distinct forms of synaptic prescriptions (learning rules). 
Here, we report a solvable network model, endowed with Hebbian synaptic
plasticity, which is able to learn either
discrete or continuous attractors, depending on the frequency of presentation
of stimuli and on the structure of sensory coding.
A continuous attractor is learned when experience matches sensory coding,
i.e. when the distribution of experienced stimuli matches the distribution of
preferred stimuli of neurons.
In that case, there is no processing of sensory information and neural activity displays
maximal entropy.
If experience goes beyond sensory coding, processing is initiated and the continuous 
attractor is destabilized into a set of discrete attractors.

\end{abstract}

\maketitle

Recurrent neural network models display persistent activity, i.e. stable attractor 
states, selective for the input stimuli (\cite{amari72},\cite{hopfield82},\cite{amit89})).
Discrete attractors are naturally generated by Hebbian plasticity
(\cite{hebb49}), storing the correlations in the input, and have been
widely used in modeling neural representations of learned objects.
Continuous attractors, on the other hand, model neural representations of
continuous variables, such as retinal angle of visual stimuli
(\cite{benyishai95}), eye position (\cite{seung96}), or space coordinates (\cite{battaglia98}).
In order to build a continuous attractor, i.e. a subspace of marginally stable
neural patterns, either homeostatic synaptic mechanisms are introduced (\cite{renart03},
\cite{blumenfeld06}), or fine tuning of parameters 
is required (\cite{tsodyks95}, \cite{koulakov02}).

Recently, a series of experiments revealed that exposure of subjects to
stimuli from a morphing sequence is able to generate, in the recorded neural
activity, discrete attractors as well as continuous ones
(\cite{freedman01},\cite{wills05},\cite{leutgeb05}).
In particular, in both \cite{wills05} and \cite{leutgeb05}, activity of
hyppocampal cells of rats is recorded while they move into boxes of
morphing shapes.
Varying the box shape, the former experiment revealed a sharp
modulation (two discrete attractors) of neural activity, while the latter showed a gradual
variation (a continuous attractor).
Moreover, the well known continuous coding of space, given by the place
fields of hyppocampal cells (\cite{okeefe78}), lays upon a discrete coding
of shapes (\cite{wills05}).
Hence, discrete and continuous attractors coexist in the same neural module, and this
raises the question of whether selection of the attractor
type depend only on the training setup, and whether the same network can
generate both, via the same synaptic plasticity mechanism.

In this letter, we present a binary neural network model, subject to Hebbian
plasticity dynamics, which can be solved analytically (a similar model has
been recently published in \cite{bernacchia07}).
Previous solvable binary models consider plasticity as exclusively driven by
the input stream, and treat recurrent neural dynamics only after learning.
In our model, instead, the recurrent dynamics plays an active role in
plasticity, giving rise to a surprisingly rich phenomenology.
For rather typical choices of the form of the input tuning curve
(\cite{dayan01}), the network autonomously develops either continuous or
discrete attractors, depending on the frequency of occurrence of stimuli.
If the distribution of presented stimuli equals the distribution of preferred
stimuli of neurons, the emergent attractor is continuous.
This corresponds to the case in which the occurrence of stimuli matches what
is expected from sensory coding, there is no processing of sensory information, and
neuronal activity displays maximal entropy.
Conversely, if subsets of stimuli are presented more frequently, a
discrete attractor appear for each highlighted set.

The model network consists of binary neurons, $S_i=\pm 1$,
and binary synapses, $J_{ij}=\pm 1$  ($i,j=1,\ldots,N$). The efficacy of a synapse connecting neurons 
$j$ to $i$ follows a stochastic Hebb's plasticity rule (\cite{tsodyks90}): if
neurons $i$ and $j$ are in correlated states ($S_i=S_j$) and $J_{ij}=-1$, then
$J_{ij}\to +1$ with probability $p$. If the neurons are in anti-correlated
states ($S_i=-S_j$) and  $J_{ij}=+1$, then $J_{ij}\to -1$ with probability
$p$.
Other configurations are unchanged.
Mean-field dynamics, describing how the synaptic efficacies vary on average
(denoted by angular brakets), is given by

\begin{equation}
\label{syndyn0}
p^{-1}\frac{d\big<J_{ij}(t)\big>}{dt}=-\big<J_{ij}(t)\big>+S_i(t)S_j(t)
\end{equation}
Synapses store the memory of past correlations in neural activity, up to a 
timescale $p^{-1}$, and regulate the recurrent currents received by neurons.
The recurrent input to neuron $i$ is equal to (\cite{willshaw69})

\begin{equation}
\label{reccur}
R_i(t)=\frac{1}{2N}\sum_{j=1}^NJ_{ij}(t)S_j(t)
\end{equation}
Beside the recurrent current, representing the ``memory trace'', each neuron
$i$ receives an external sensory signal, determined by its "preferred
stimulus" $\eta_i$.
The external current to neuron $i$, upon presentation of stimulus
$\alpha$, is defined by the ``tuning curve'' $E(\eta_i-\alpha)$, describing
how the external current is modulated by changes in the presented stimulus.
We assume that the tuning curve is
monotonically increasing ($E'>0$) and centered ($E(0)=0$), and we define
the space of stimuli as the unitary segment,
$T=(-\frac{1}{2},+\frac{1}{2})$ ($\eta_i,\alpha\in T$).
However, results can be generalized to a decreasing as well as to a periodic
tuning curve (to be presented elsewhere).
The preferred stimulus $\eta_i$ is randomly assigned to each neuron, following
a distribution $\omega(\eta)$.

The total current afferent to each neuron is the sum of the recurrent and
the external currents, and neural activity 
is $+1$ or $-1$, respectively, if the total current is above or below
threshold (set to zero), i.e. 
(\cite{amari72}, \cite{hopfield82})

\begin{equation}
\label{neudyn0}
S_i(t+dt)=\mbox{sign}\Big[R_i(t)+E(\eta_i-\alpha)\Big]
\end{equation}
We present a theoretical analysis of the network dynamics, in the limit of an
infinite number of neurons, and then we show computer simulations to corroborate the emerging picture.

It turns out that the dynamics of neural activity can be described by a label
$\mu(t)\in T$, such that

\begin{equation}
\label{mupatt}
S_i(t)=\mbox{sign}\Big(\eta_i-\mu(t)\Big)
\end{equation}
Hence, a single number $\mu$, at each time step, defines the entire
neural activity pattern. 
Eq.(\ref{mupatt}) holds trivially if the recurrent currents are negligible
with respect to the external currents, and the activity label is just equal to
the presented stimulus, $\mu=\alpha$. 
However, a self-consistent argument demonstrates that, under
quite general conditions, Eq.(\ref{mupatt}) holds, for some $\mu$, even for
large recurrent currents, after the synapses have
``learned'' the space of stimuli $T$. 
We know, from Eq.(\ref{syndyn0}), that the average synaptic matrix
$\big<J_{ij}\big>$ is a linear functional of
the activity product $S_i(t)S_j(t)$.
Then, since neural activities are mapped
onto the space of stimuli ($\mu\in T$ in Eq.(\ref{mupatt})), the synaptic matrix can be written as

\begin{equation}
\label{intsyn}
\big<J_{ij}(t)\big>=\int_T  \psi(\mu,t)\;\mbox{sign}(\eta_i-\mu)\;\mbox{sign}(\eta_j-\mu)\;d\mu
\end{equation}
where $\psi(\mu,t)$ is the ``distribution of stored patterns'', expressing the
relative weight of different patterns, labeled by $\mu$, in the synaptic memory matrix. 
It is positive ($\psi(\mu,t)>0$) and normalized ($\int_T\psi(\mu,t)d\mu=1$).
In the following, we study the stationary behaviour of $\mu$ and
$\psi$, describing respectively the neural and synaptic states, and
we discuss their physical interpretations.

\begin{figure}
\includegraphics[scale=0.5]{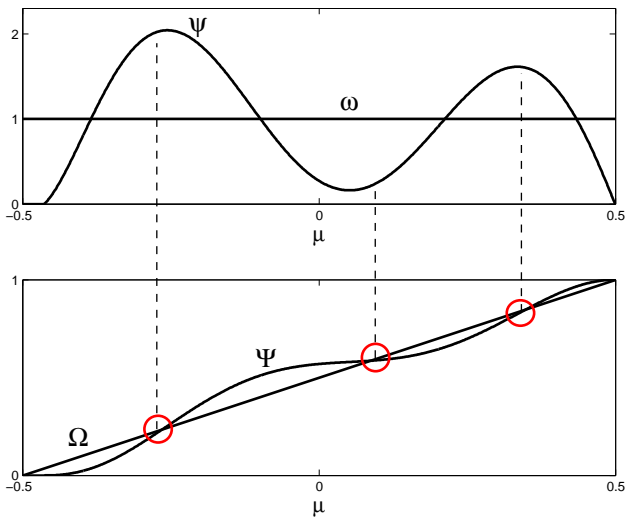}
\caption{An example of distribution of stored patterns and of preferred stimuli, the density (top, respectively 
$\psi$ and $\omega$), and the cumulative distribution functions (bottom, $\Psi$ and $\Omega$).
The three intersections of $\Psi$ and $\Omega$ (red circles) give
approximately the stationary states of recurrent neural dynamics, one unstable
(middle) and two stable (left and right).
}
\label{example}
\end{figure}

We assume that the dynamics of neurons is much faster than that of synapses,
which are essentially frozen on the neural timescale. 
Upon presentation of a given stimulus $\alpha$, neural activity reaches
immediately a stationary state $\mu_s$, resulting from the competition of
external and recurrent currents.
Then, synaptic learning applies over the state $\mu_s$, until a new stimulus
is presented and a new state $\mu_s$ is reached.
By using Eqs.(\ref{reccur}),(\ref{neudyn0}),(\ref{mupatt}),(\ref{intsyn}) with
$\mu(t+dt)=\mu(t)=\mu_s$, and averaging over preferred stimuli, the equation
for $\mu_s$ is found, 

\begin{equation}
\label{neustat}
E(\mu_s-\alpha)+\Psi(\mu_s)-\Omega(\mu_s)=const
\end{equation}
where $\Psi$ and $\Omega$ are the cumulative distribution functions of,
respectively, the density distributions of the stored patterns $\psi$ and
of the preferred stimuli $\omega$ (an example is given in Fig.\ref{example}), 
i.e. $\Psi(\mu)=\int_{-1/2}^\mu\psi(\mu')d\mu'$, and $\Omega(\eta)=\int_{-1/2}^\eta
\omega(\eta')d\eta'$, and
$const=\int_0^1(\Psi-\Omega)\;d(\Psi+\Omega)/2$ 
(since synapses are frozen, the time dependence of $\psi$ is omitted). 
The stationary state $\mu_s$ is stable when
$E'(\mu_s-\alpha)+\psi(\mu_s)-\omega(\mu_s)>0$.

The neural dynamics is a trade-off between the external drive and the recurrent dynamics.
In order to get a physical intuition of Eq.(\ref{neustat}), we consider
separately the recurrent and external contributions: the general solution will
be somewhere in between the two cases.
If the external input is strong, Eq.(\ref{neustat}) reduces to
$E(\mu_s-\alpha)=0$, whose solution is $\mu_s=\alpha$. In that case, the
strong input forces the neural pattern to match the stimulus.
On the other hand, when the recurrent contribution dominates, Eq.(\ref{neustat}) becomes
$\Psi(\mu_s)-\Omega(\mu_s)=const$.
The solutions of this equation are the attractors of the recurrent
neural dynamics, in absence of the stimulus.
If the distribution of stored patterns matches the distribution of preferred stimuli,
i.e. if $\Psi=\Omega$ everywhere, then $const=0$, and all values of $\mu_s\in
T$ are solutions, 
all marginally stable: this corresponds to a continuous attractor.
In general, if $\Psi\neq\Omega$, discrete attractors emerge.
An illustrative example is given in Fig.\ref{example}, where $const$ is
neglected: two discrete attractors appear near the maxima of $\psi$.
A continuous attactor would appear when the two curves superimpose, intersecating at all points.

The slow dynamics of $\psi$ depend on the sequence of input
stimuli $\alpha$, through the sequence of the resulting stationary neural patterns $\mu_s$.
We assume that each presentation of a stimulus $\alpha$ is drawn
independently from a distribution $\phi(\alpha)$. 
Starting from the synaptic mean-field dynamics, Eq.(\ref{syndyn0}), using
Eqs.(\ref{mupatt}),(\ref{intsyn}) and averaging over the presented stimuli, the dynamics of $\psi$ is written as

\begin{equation}
\label{syndyn}
p^{-1}\frac{\partial\psi}{\partial t}\Big|_{\mu,t}=-\psi(\mu,t)+\int\phi(\alpha)\delta\big(\mu-\mu_s\big)d\alpha
\end{equation}
where $\delta$ is the Dirac pulse function, and $\mu_s$ depends, via Eq.(\ref{neustat}), on the
stimulus $\alpha$ and on the distribution $\psi$ itself.
Eq.(\ref{syndyn}) implies that, from stimulus to stimulus,
$\psi$ strenghtens the memory of each activated pattern of neural activity
$\mu_s$, and slightly weakens all the others.
The external-recurrent trade-off play a crucial role in determining the
stationary solution $\psi_s(\mu)$ of Eq.(\ref{syndyn}): when external currents dominate,
$\mu_s=\alpha$, and the stationary solution simply
replicates the distribution of presented stimuli, i.e. $\psi_s=\phi$.
In that case, the synapses store exactly the presented input patterns, weighted by their 
relative frequency of appearance.
Then, a continuous attractor ($\psi=\omega$) forms if the distribution of
presented stimuli equals the distribution of preferred stimuli of neurons, i.e. if
$\phi=\omega$.
Surprisingly, a continuous attractor stabilizes even if recurrent currents give a finite contribution, 
as demonstrated in the following. 

In general, a stationary solution of Eq.(\ref{syndyn}) is difficult to find.
Neverthless, we can calculate self-consistently $\psi_s(\mu)$ for stored patterns $\mu$ which correspond to stationary states of
the recurrent neural dynamics, i.e. those satisfyng $\Psi_s(\mu)-\Omega(\mu)=const$.
The solution is

\begin{equation}
\label{fixsyn}
\psi_s(\mu)=\phi(\mu)\bigg[\frac{E'(0)-\omega(\mu)}{E'(0)-\phi(\mu)}\bigg]
\end{equation}
if positive, zero otherwise.
This solution is stable if $E'(0)>\phi(\mu)$.
The contribution of the recurrent dynamics to plasticity is given by the term in
square brakets, which disappears in the limit of large external input
($E'(0)\rightarrow+\infty$), for which we recover $\psi_s=\phi$.

We stress that the solution (\ref{fixsyn}) is not valid for all values of
$\mu$, but only at the fixed points of $\Psi_s(\mu)-\Omega(\mu)=const$.
However, if $\phi=\omega$ the solution of Eq.(\ref{fixsyn}) is $\psi_s=\omega$,
that corresponds to a continuous attractor, for which all points are fixed,
and the solution (\ref{fixsyn}) is valid everywhere.
Hence, if the distribution of presented stimuli equals the distribution of
preferred stimuli of neurons, a continuous attractor may stabilize, as a result
of the combined synaptic and neural dynamics.
The continuous attractor, in terms of plasticity dynamics, is stable if
$E'(0)>$Max$\big(\omega(\mu)\big)$, i.e. is favored by a strong external input and
a homogeneous distribution of preferred stimuli.

If $\phi\neq\omega$, only discrete attractors stabilize,
and we can calculate the stationary solution $\psi_s$ only at those points.
However, we expect Eq.(\ref{fixsyn}) to be a good approximation in the
vicinity of a continuous attractor ($\phi\simeq\omega$), if it is stable and
positive ($E'(0)>\phi, \omega$).
In that case, patterns are stored more efficiently if the corresponding
stimuli are more experienced and less preferred by neurons (large $\phi$ and small $\omega$).
If the external input is weak ($E'(0)<\phi(\mu)$), 
the solution (\ref{fixsyn}) is unstable: a continuous attractor is prohibited, while discrete
attractors are still allowed, whose $\psi_s(\mu)$ diverges.
In that case, synaptic changes are driven by the recurrent synaptic structure
itself, rather than by the external signal, and we expect that the initial synaptic
structure strongly affects the dynamics.

In order to check to what extent the network processes sensorial information,
we calculate the entropy of the distribution of the total neural activity,
$A=\frac{1}{2N}\sum_iS_i$, which is a continuous variable in the limit of
large $N$, and whose distribution
is denoted by $\rho(A)$.
In stationary conditions, using Eq.(\ref{mupatt}) and averaging over preferred
stimuli, the entropy is calculated as

\begin{equation}
\label{entract}
H=-\int_{-\frac{1}{2}}^\frac{1}{2} \rho(A)\log\rho(A)dA=\int_0^1\mbox{log}\Big(\frac{d\Omega}{d\Psi}\Big)d\Psi
\end{equation}
which is negative, and corresponds to (minus) the information gained when using 
a code based on the distribution $\Psi$ rather than on $\Omega$ (if $\Psi$ is the true 
distribution).
If the two distributions are equal, the network is in a continuous attractor
state, and the entropy of neural activity is maximal (zero, and $A$ is
uniformly distributed, $\rho=1$).
Conversely, if $\Psi$ is different from $\Omega$, the information inside neural activity 
equals the information gained by using the synaptic matrix, structured by learning, instead of 
sensory coding.
Once normalized, Eq.(\ref{fixsyn}) is taken as an approximation of the
stationary solution $\psi_s$, whose cumulative $\Psi_s$ is used for computing the theoretical entropy of Eq. (\ref{entract}).

We illustrate our predictions by $100$ computer simulations of the dynamics of the network
(Eqs.(\ref{syndyn0}),(\ref{reccur}) and (\ref{neudyn0})), which is composed of $1000$ neurons. 
For each simulation, synapses and neurons are initialized at random, and distribution functions 
of preferred ($\omega$) and presented ($\phi$) stimuli are chosen randomly from the set of fourier 
series truncated at wave number 5 
(i.e. $1+\sum_{i=1}^5[a_i\sin(\frac{i\mu}{2\pi})+b_i\cos(\frac{i\mu}{2\pi})]$, where $a_i,b_i$ are random
in the interval $(-0.5,0.5)$, and only positive series are taken).
Then, $1000$ preferred stimuli are drawn from $\omega$, for the $1000$ neurons, and $2000$ 
stimuli, to be presented, are drawn from $\phi$.
The tuning curve is linear, and its angular coefficient $E'$ is chosen, for each 
simulation, from a uniform 
distribution (in the interval $(0,5)$), to which is added the maximum of $\phi$ and $\omega$
(then (\ref{fixsyn}) is stable and positive).  
Each stimulus is presented for $20$ time steps, and the learning 
timescale is set $p=10^{-4}$.
In order to analise stationary conditions, neural activity is recorded only after presentation 
of the first $1000$ stimuli, and entropy is evaluated by the neural activity (sampled in $50$ bins)
at the last step of each of the remaining stimuli, from $1001$ to $2000$. 

\begin{figure}
\includegraphics[scale=0.5]{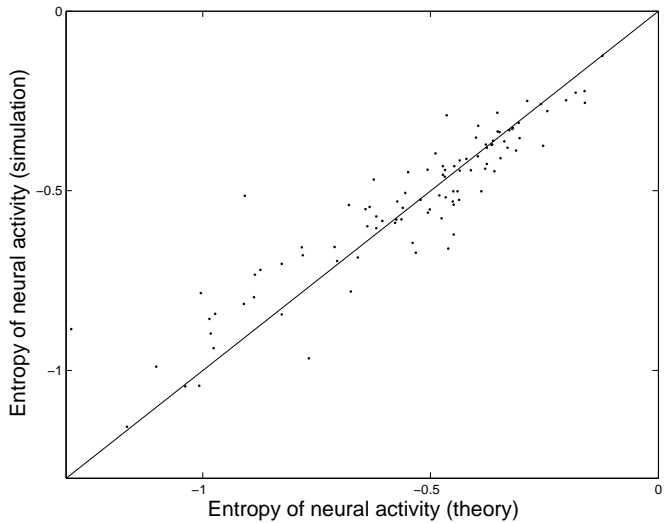}
\caption{Entropy of neural activity, measured in simulations, plotted
against the theoretical prediction. 
Each point is one of the $100$ simulations of the network dynamics.}
\label{theosim}
\end{figure}

Theoretical predictions are in good agreement with the simulation results, as shown in Fig.\ref{theosim}, 
where the measured entropy of neural activity is plotted against the
prediction of Eq.(\ref{entract}), each point is one simulation.
The agreement is especially good for large entropies, where the network is close to a continuous 
attractor state, and for which Eq.(\ref{fixsyn}) is a better approximation.
The observed contribution of recurrent dynamics to plasticity is an average
$65\%$ decrease in the entropy, with respect to the case of large external
drive, separately studied. 

In summary, we have shown that a simple plastic network can generate
both discrete and continuous attractor states, depending on the
statistics of experienced stimuli and the structure of sensory coding.
The network provides a candidate explanation for the apparent discrepancy between the two experiments
\cite{wills05} and \cite{leutgeb05}, in which similar experimental conditions
result in the observation of the two different types of attractors.
Here, the continuous attractor has been shown to be more stable when the
distribution of preferred stimuli is homogeneuos, such as commonly observed in the brain.
We leave for a future work the study of the network finite-size effects,
eventually wasting a perfect continuous attractor state.
From the computational point of view, the network introduces a novel
functionality: the network divides stimuli in clusters (discrete attractors)
when subsets of stimuli occur more frequently than what is expected from sensory coding.
In that case, neural activity expresses exactly the information gained by the 
learned synaptic matrix. 
If the statistics of presented and preferred stimuli match, neural activity
displays maximal entropy (no additional information), and the network abandons any tentative to 
recognize clusters of stimuli, but it still provides their representation via the continuous attractor.
The present network represents an effort to bridge high-level (memory) areas of the
brain, which could be modeled by recurrent associative networks
(\cite{amit89}), and early sensory brain areas, in which continuously varying
stimuli are represented by smooth tuning curves (\cite{dayan01}).
Here, the recurrent contribution to plasticity was found to decrease substantially
the entropy of neural activity.

This work was supported by the european E2-C2 grant. The author would like to
 thank Daniel Amit, Sandro Romani and Stefano Fusi for valuable discussions.

\end{document}